\documentclass[twocolumn,showpacs,preprintnumbers,amsmath,amssymb]{revtex4}
\usepackage{dcolumn}
\usepackage{bm}
\usepackage[dvips]{graphicx}
\usepackage{wrapfig,subfigure}
\usepackage{amssymb}
\usepackage{color}

\newcommand{\tr}{t_{\rm r}}
\newcommand{\qb}{{\bf q}}

\newcommand{\fb}{{\bf f}}
\newcommand{\cb}{{\bm \chi}}
\newcommand{\xib}{{\bm \xi}}

\newcommand{\qbo}{{\bf q}_{\rm opt}}

\newcommand{\fbo}{{\bf f}_{\rm opt}}

\newcommand{\kb}{{\bf k}}
\newcommand{\kbo}{{\bf k}_{\rm opt}}
\def\<{\lesssim}
\def\>{\gtrsim}

\begin{document}

\title{Thermally activated switching in the presence of non-Gaussian noise}

\author{Lora Billings$^{(1)}$, Mark I. Dykman$^{(2)*}$,  and Ira B. Schwartz$^{(3)}$}
\email{dykman@pa.msu.edu}
\affiliation{$^{(1)}$Department of Mathematical Sciences, Montclair State University, Montclair, NJ 07043\\
$^{(2)}$ Department of Physics and Astronomy, Michigan State University, East Lansing, MI 48824\\
$^{(3)}$ US Naval Research Laboratory, Code 6792, Nonlinear System Dynamics Section, Plasma Physics Division, Washington, DC 20375}
\date{\today}

\begin{abstract}
We study the effect of a non-Gaussian noise on interstate switching activated primarily by Gaussian noise. Even weak non-Gaussian noise can strongly change the switching rate. The effect is determined by all moments of the noise distribution. The explicit analytical results are compared with the results of simulations for an overdamped system driven by white Gaussian noise and a Poisson noise. Switching induced by a purely Poisson noise is also discussed.
\end{abstract}

\pacs{05.40.-a, 72.70.+m, 05.70.Ln, 05.40.Ca }
\maketitle

Much progress has been made recently in the studies of
switching between coexisting stable states, primarily because switching can be now
investigated for a large variety of well-controlled micro- and mesoscopic systems ranging from trapped electrons and atoms to
Josephson junctions and to nano- and micro-mechanical oscillators
\cite{Lapidus1999,Siddiqi2005,Aldridge2005,Kim2005,Gommers2005,Stambaugh2006,Abdo2007,Lupascu2007,Katz2007,Serban2007}. Fluctuations
in these systems are usually due to thermal or externally applied Gaussian
noise. However, as the systems become smaller, an increasingly important role
may be played also by non-Gaussian noise. It may come, for example,
from one or a few two-state fluctuators hopping at random between the states, in which case
the noise may be often described as a telegraph noise.

The switching probability is sensitive to a non-Gaussian noise. This
sensitivity attracted much attention after it was proposed \cite{Tobiska2004}
to use switching in Josephson junctions to measure the full counting
statistics in electronic circuits \cite{Levitov1992,Nazarov2007}. Several
theoretical \cite{Pekola2004,Ankerhold2007a,Sukhorukov2007,Grabert2008} and
experimental \cite{Timofeev2007,Huard2007} papers on measuring the 3rd moment
of the current distribution from the switching rates were published recently,
and different theoretical approaches were compared in
Refs.~\onlinecite{Novotny2008,Sukhorukov2008}.

In this paper we study switching induced by Gaussian noise in the presence of
an additional non-Gaussian noise. Even where the latter has a
smaller intensity than the Gaussian noise, its effect on the
switching rate may be exponentially strong. We show that it can be described in a simple form in terms
of the noise characteristic functional, thus accounting for {\it all} moments of
the noise distribution. The
analytical results are compared with simulations for an overdamped system
driven by white Gaussian noise and a Poisson noise. We also consider switching induced by a Poisson noise alone; here, the result for the rate may be qualitatively different from that in the weakly non-Gaussian noise approximation.

The potentially strong effect of an extra modulation, whether random or regular, on the rate of Gaussian
noise induced switching can be understood from the well-established
picture of the switching dynamics. Switching events result from large rare
noise outbursts. For Gaussian noise, the switching rate is $W\propto
\exp(-R/D)$, where $R$ is the activation energy and $D$ is the noise
intensity \cite{Freidlin_book}. Even though switching happens at random,
the system trajectories followed in switching form a narrow tube in the space
of dynamical variables $\qb=(q_1,q_2,\ldots)$
centered at the most probable (optimal) switching path  $\qbo(t)$.

One can think of the effect of an additional modulation in terms of a
generalized work done by the modulation on the system moving along $\qbo(t)$
\cite{Smelyanskiy1997c,Dykman1997}. This work changes the activation
barrier. The change $\delta R$ is proportional to the modulation
amplitude. Therefore the overall change of the switching rate
$\propto\exp(-\delta R/D)$ depends on the modulation amplitude
exponentially. The switching rate gives the probability current from the
occupied state \cite{Kramers1940}, making it  an observable
quantity. Thus, if
the modulation is random,
one has to simply average the factor $\exp(-\delta R/D)$ over realizations of the
modulation.

Since $\delta R$ is linear in the characteristic {\it amplitude}
of random modulation, the ratio $\delta R/D$ does not have to be small even
where the modulation {\it intensity}, which is quadratic in the amplitude, is
smaller than $D$. However, the distribution of non-Gaussian modulation may
decay slower than Gaussian on the tail. To determine whether the effect of a non-Gaussian noise on switching may be regarded as a perturbation one has to
compare the probabilities of appropriate large fluctuations induced by the
Gaussian and non-Gaussian noises, taking into account {\it all} moments of the distribution.

We study switching for a system described by the Langevin equation
\begin{equation}
\label{eq:eom}
\dot\qb={\bf K}(\qb)+\fb(t)+\xib(t).
\end{equation}
We assume that, in the absence of noise, the system has a stable stationary state $\qb_A$ and a saddle point $\qb_{\cal S}$ on the boundary of the basin of attraction to $\qb_A$, with  ${\bf K}(\qb_A)={\bf K}(\qb_{\cal S})=0$. Switching from the stable state is due to the forces $\fb(t)$ and $\xib(t)$, which are the Gaussian and non-Gaussian noises, respectively. We separate them, since physically they often come from different sources. It is convenient to characterize $\fb(t)$ by its probability density functional  ${\cal P}_{\fb}[\fb(t)]= \exp\left(-{\cal R}_{\fb}/D\right)$,
\begin{equation}
\label{eq:Gauss_Functional}
{\cal R}_{\fb}[\fb(t)]= \frac{1}{4}\int dt\,dt'\,\fb(t)\hat{{\cal F}}(t-t')\fb(t'),
\end{equation}
where $\hat{{\cal F}}(t-t')/2D$ is the inverse of the pair correlator of $\fb(t)$. The characteristic noise intensity $D$ is small, so that the switching rate $W\ll \tr^{-1}, t_{\rm c}^{-1}$, where $\tr$ is the relaxation time of the system and $t_{\rm c}$ is the noise correlation time.
The non-Gaussian noise is more conveniently described for our purpose by the characteristic functional
\begin{equation}
\label{eq:char_functional_definition}
\tilde{\cal P}_{\xib}[\kb]=\left\langle\exp\left[i\int dt\kb(t)\xib(t)\right]\right\rangle_{\xib},
\end{equation}
where $\langle\ldots\rangle_{\xib}$ means averaging over $\xib(t)$.

We first consider the case where the intensity of the non-Gaussian noise $\xib$ is smaller than $D$. We will disregard corrections proportional to this intensity, but the ratio of the characteristic amplitude $g_0$ of $\xib$ to $D$ will not be assumed small. The switching rate can be written as
\begin{eqnarray}
\label{eq:W_general_w_nonGauss}
&&W=C\left\langle\exp[-{\cal R}[\xib]/D]\right\rangle_{\xib},\\ &&{\cal R}[\xib]
=\min \left\{ {\cal R}_{\fb}+i\int dt\,\kb(t)\left[\dot \qb - {\bf K}-\fb(t)-\xib(t)\right]\right\},\nonumber
\end{eqnarray}
where $C$ is the prefactor that weakly depends on the noise intensity. The minimum is taken over trajectories $\fb(t), \qb(t), \kb(t)$ that satisfy boundary conditions $\fb (t), \kb(t)\to 0$ for $t\to\pm\infty$, $\qb_{t\to-\infty}\to \qb_A, \qb_{t\to\infty}\to \qb_{\cal S}$. This formulation was proposed in the weak-noise limit \cite{Dykman1990,Dykman1998b} for a time-periodic $\xib(t)$, in which case $\qb_{A,{\cal S}}$ are also periodic and there is no averaging over $\xib$. The variational problem (\ref{eq:W_general_w_nonGauss}) describes coupled optimal trajectories $\fbo(t), \qbo(t), \kbo(t)$, with
$\fbo(t)$ being the most probable noise realization that brings the system to
the saddle on the basin boundary of the initially
occupied state.

It is known from variational calculus that, to first order in $\xib$, the
effect of $\xib(t)$ on ${\cal R}$ can be calculated along the optimal
trajectory unperturbed by $\xib(t)$. Such a trajectory is an instanton. Its
typical duration is $\sim \max(\tr, t_{\rm c})$. It is translation-invariant with respect to time and can be centered at any
time $t_0$.
If $\xib(t)$ is periodic, it lifts time-translation symmetry and fixes $t_0$ (modulo the period)  so as to
minimize ${\cal R}[\xib]$.

If $\xib(t)$ is a stationary noise, the switching rate $W$ is independent of
time. In this case one can think not of the adjustment of the
instanton center $t_0$ to $\xib(t)$, but, equivalently, of the adjustment of
$\xib(t)$ to $t_0$ so as to maximize the overall probability of
switching. This adjustment provides the major contribution to the value of $W$
when the averaging over realizations of $\xib(t)$ is performed in
Eq.~(\ref{eq:W_general_w_nonGauss}) using a solution with a given $t_0$.

From Eqs.~(\ref{eq:char_functional_definition}), (\ref{eq:W_general_w_nonGauss}) one obtains a simple expression for the switching rate
\begin{eqnarray}
\label{eq:W_general_answer}
W=W^{(0)}A_{\rm sw},\qquad A_{\rm sw}=\tilde{\cal P}_{\xib}[i\cb/D],
\end{eqnarray}
where $W^{(0)}$ is the switching rate in the absence of non-Gaussian
noise. The factor $A_{\rm sw}$ describes the effect of non-Gaussian noise. It is expressed in a closed form in terms of the noise
characteristic functional calculated for function $\cb(t)=-i\kbo^{(0)}(t)$,
where $\kbo^{(0)}(t)$ is the solution of the variational problem
(\ref{eq:W_general_w_nonGauss}) for $\xib={\bf 0}$. The real function $\cb(t)$ is the logarithmic susceptibility which describes the
linear response of $\log W$ to a perturbation
\cite{Smelyanskiy1997c,Dykman1997,Assaf2008}. The structure of
Eq.~(\ref{eq:W_general_answer}) resembles that of the expression for a large
fluctuation probability in a birth-death system with non-Gaussian modulation of reaction rates \cite{Dykman2008}.

From Eq.~(\ref{eq:W_general_answer}), the effect of a non-Gaussian noise on the switching rate is determined by the ratio of the noise amplitude to the Gaussian noise intensity $D$. Equation~(\ref{eq:W_general_answer}) applies to both underdamped and overdamped systems. Examples of calculating $\kbo^{(0)}$ can be found in Refs.~\onlinecite{Smelyanskiy1997c,Dykman1997,Assaf2008} and papers cited therein.

As an illustration we will consider the case of a one-component $\delta$-correlated Poisson noise $\xi(t)$ with pulse area $ g$ and mean frequency $\nu$. Using the explicit form of the noise characteristic functional \cite{FeynmanQM}, we obtain
\begin{eqnarray}
\label{eq:Asw_for_1D_Poisson}
A_{\rm sw}=\exp\left\{-\nu\int dt\,\left[1-\exp\left(-\chi(t) g/D\right)\right]\right\},
\end{eqnarray}
where $\chi(t)$ is the corresponding component of the logarithmic susceptibility. If $ g/D\ll 1$, $\log A_{\rm sw}$ is a series in $ g/D$. The coefficients in this series describe the effects of the moments of the Poisson noise on the switching rate. In the opposite case, $ g/D \gg 1$ (but the Poisson noise intensity $\nu g^2\ll D$), if $\chi(t)$ becomes negative, then
$\log A_{\rm sw}\approx \nu \left[2\pi D/ g\ddot \chi(t_m)\right]^{1/2}\exp[-\chi(t_m) g/D]$
where $t_m$ is the instant where $-\chi(t)$ is maximal. If $\chi(t)\geq 0$ for all $t$ and $ g/D \gg 1$, the major contribution to $A_{\rm sw}$ comes from the region of small $\chi(t)$. If $\chi(t)$ is small only for $|t|\to\infty$, where it decays exponentially with $|t|$, then $\log A_{\rm sw}\propto \nu\tr\log( g/D)$, to leading order in $ g/D$.

The Poisson noise distribution does not fall off as steeply as Gaussian. This imposes a limitation on the range of $g/D$ where Poisson noise may be treated as a perturbation and the above theory applies. To see the far-tail effect we consider switching due to a purely Poisson noise, where $\fb ={\bf 0}$ in equation of motion (\ref{eq:eom}). We will use the method of optimal fluctuation, as for some other types of non-Gaussian noise \cite{Mckane1989}.

The switching rate is determined, to logarithmic accuracy, by the integral over trajectories $\kb(t),\qb(t)$ of the functional
$\left\langle\exp\left\{-i\int dt \kb(t)[\dot \qb -{\bf K}-\xib(t)]\right\}\right\rangle_{\xib}$
\cite{Luciani1987}. We assume for brevity that different components of the Poisson noise are independent short pulses with areas ${\bf g}=( g_1, g_2,\ldots)$ and with average frequency $\nu$. It is convenient to consider a zero-mean noise, $\xib(t) \to \xib(t) - \nu{\bf g}, \;{\bf K} \to {\bf K} + \nu{\bf g}$; we assume that $\qb_A,\qb_{\cal S}$ are also appropriately shifted. Of interest for switching are trajectories that approach the saddle point \cite{Dykman1990}. In the spirit of the method of optimal fluctuation, for small $| {\bf g}|$ and for $\nu\lesssim \tr^{-1}$ the integral over trajectories $\qb(t),\kb(t)$ can be calculated by steepest descent. This gives
\begin{eqnarray}
\label{eq:W_pure_Poisson_General}
&&W=C'\exp[-R_P], \quad R_P=\min\int dt \left(i\kb\dot\qb - H\right),\\
&&H\equiv H(\qb,i\kb)=-\nu \left(1+i{\bf g}\kb-e^{i{\bf g}\kb}\right) + i\kb {\bf K}(\qb).\nonumber
\end{eqnarray}
The variational problem (\ref{eq:W_pure_Poisson_General}) determines the optimal switching trajectory $\qbo(t),\kbo(t)$. It starts at $t\to -\infty$ at $\qb\to\qb_A, \kb\to {\bf 0}$ and goes to $\qb\to\qb_{\cal S}, \kb\to {\bf 0}$ for $t\to\infty$. On this trajectory $H=0$. As in systems driven by white Gaussian noise \cite{Freidlin_book}, the optimal trajectory can be described as a Hamiltonian trajectory of an auxiliary system with coordinate $\qb$, momentum $i\kb$, and Hamiltonian $H$. A similar approach was proposed in Ref.~\onlinecite{Sukhorukov2007}. However, the characteristic functional was not specified and the explicit analysis took into account only the 3rd moment of the noise distribution, which was considered a perturbation; therefore the results do not describe switching due to large Poisson fluctuations.

The switching exponent in Eq.~(\ref{eq:W_pure_Poisson_General}) is $R_P\gg 1$ for small $|{\bf g}|$. However, in contrast to the case of Gaussian noise, $R_P$ is not proportional to the reciprocal noise intensity $\nu{\bf g}^2$. Nor does it scale like reciprocal noise amplitude $|{\bf g}|^{-1}$, although $R_P|{\bf g}|$ appears to slowly vary with $|{\bf g}|$.

An explicit dependence of the switching rate on the intensity $g$ of Poisson-distributed pulses can be found for a one-variable overdamped system with equation of motion
\begin{equation}
\label{eq:eom_scalar_Poisson}
\dot q=-U'(q) + f(t) +\xi(t).
\end{equation}
Here, $U(q)$ is the effective potential. The stationary states $q_A$ and $q_{\cal S}$ correspond to the minimum and the barrier top of $U(q)$.

If $f(t)$ is white Gaussian noise, $\langle f(t)f(t')\rangle = 2D\delta (t-t')$, and the Poisson noise is weak, the Poisson-noise induced factor in the switching rate $A_{\rm sw}$ is described by Eq.~(\ref{eq:Asw_for_1D_Poisson}) with $\chi(t)=-f_{\rm opt}^{(0)}(t)/2=-\dot q_{\rm opt}^{(0)}(t)$ and with $\dot q_{\rm opt}^{(0)}=U'\left(q_{\rm opt}^{(0)}\right)$.

In the opposite case where switching is due to purely Poisson noise, i.e., $f=0$ in Eq.~(\ref{eq:eom_scalar_Poisson}), from Eq.~(\ref{eq:W_pure_Poisson_General})
\begin{eqnarray}
\label{eq:switching_exponent_1D_Poisson}
R_P=\frac{1}{g}\int\nolimits_{\tilde q_A}^{\tilde q_{\cal S}} dq \kappa(q),
\quad \kappa=\log\left\{1+\left[\kappa U'(q)/g\nu\right]\right\}.
\end{eqnarray}
Here, $\tilde q_A$ and $\tilde q_{\cal S}$ are the shifted extrema of the potential given by equation $U'(q)=g\nu$. From Eq.~(\ref{eq:switching_exponent_1D_Poisson}), $R_P\sim r_P\log (r_P/\nu\tr)$, with $r_P=(\tilde q_{\cal S}-\tilde q_A)/g$.

A qualitative feature of unipolar (pulses of one sign) Poisson noise is that, for an overdamped system, it causes switching only provided the noise pulses push the system from the stable state towards the saddle. In this case  $r_P>0$. There is no switching for pulses of the opposite sign. The ``one-sidedness" of fluctuations in overdamped systems has other manifestations, which includes the work fluctuation distribution \cite{Cohen_private}. On the other hand, we expect that an underdamped system should be able to switch for Poisson pulses of any sign,
in which case there should be a critical value of damping for which switching from the state is possible for a given sign of $g$.
Equations (\ref{eq:W_pure_Poisson_General}), (\ref{eq:switching_exponent_1D_Poisson}) apply if $r_P, r_P/\nu\tr \gg 1$.

As a cause of switching, a Poisson noise is effectively weaker than a Gaussian noise if the switching exponent $R_P$ is larger than the switching exponent for the Gaussian noise. For white Gaussian noise of intensity $D$ in Eq.~(\ref{eq:eom_scalar_Poisson}), the switching exponent is $\Delta U/D$ with $\Delta U=U(q_{\cal S})- U(q_A)$ \cite{Kramers1940}. The condition $R_P > \Delta U/D$ effectively limits the range of applicability of Eq.~(\ref{eq:Asw_for_1D_Poisson}). We saw that $\log A_{\rm sw}$ becomes large provided $-\chi(t_m) g/D \gg 1$. An order of magnitude estimate shows that $-\chi(t_m) g/D\sim (\Delta U/D)R_P^{-1}\log(r_P/\nu\tr)$, and therefore from Eq.~(\ref{eq:switching_exponent_1D_Poisson}) the large $\log A_{\rm sw}$ asymptotics applies only provided $\log(r_P/\nu\tr)\gg 1$. This condition is compatible with $-\chi(t_m) g/D\gg 1$ only for very small $D$.

We now apply the above results to an overdamped system (\ref{eq:eom_scalar_Poisson}) with a double-well potential
\begin{equation}
\label{eq:quartic_potential}
U(q)=-q^2/2 + q^4/4,
\end{equation}
which has been extensively studied in the context of white-noise driven systems. In the absence of Poisson noise, the escape rate in this case is $W^{(0)}=(\sqrt{2}/\pi)\exp(-1/4D)$, and the logarithmic susceptibility for escape from the negative-$q$ well ($q_A=-1$) is $\chi(t)=-\exp(t/2)(2\cosh t)^{-3/2}$ ($\chi(t)$ has opposite sign for switching from $q_A=1$).

\begin{figure}[h]
\includegraphics[width=2.8in]{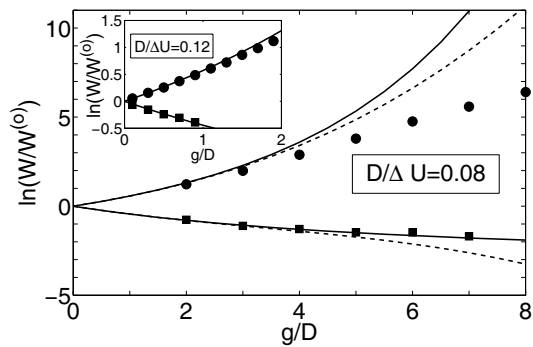}
\caption{Poisson noise induced change of the switching exponent for an overdamped Brownian particle in potential (\ref{eq:quartic_potential}). The dimensionless Poisson noise frequency is $\nu=0.5$. The data of numerical simulations are shown with circles and squares for the cases where in escape the particle moves along and opposite to Poisson pulses, respectively. The solid curves show the weak-Poisson noise theory (\ref{eq:Asw_for_1D_Poisson}) for these cases, and the dashed curves show the approximation where only the three moments of Poisson noise are taken into account.}
\label{fig:Gauss_and_Poisson}
\end{figure}

In Fig.~\ref{fig:Gauss_and_Poisson} we present results of Monte Carlo simulations of switching of an overdamped Brownian particle described by Eqs.~(\ref{eq:eom_scalar_Poisson}), (\ref{eq:quartic_potential}). They are compared with the weak Poisson noise prediction, Eq.~(\ref{eq:Asw_for_1D_Poisson}), and with the approximation where only terms up to $g^3$ are kept in Eq.~(\ref{eq:Asw_for_1D_Poisson}). The Poisson noise intensity $\nu g^2 <D$ in the whole range of studied $g/D$. For $g/D\lesssim 1$ the effect of Poisson noise is small and comes, primarily, to the change of the activation barrier, $\Delta U\to \Delta U\pm \nu g$, and the effective noise intensity, $D\to D+\nu g^2$. For larger $g/D \gtrsim 3$ the switching exponent changes significantly, as expected.

For switching in the direction opposite to Poisson noise pulses, where $\log W/W^{(0)}<0$, the numerics agrees well with Eq.~(\ref{eq:Asw_for_1D_Poisson}). This is to be expected, since the far tail of Poisson noise distribution is immaterial here; for $g/D\gtrsim 5$ the results differ noticeably from the three-moments approximation. For switching along the pulses, because of the far-tail effect, with increasing $g$ Poisson noise quickly becomes as important as white noise for chosen $D$. Therefore the perturbation theory fails and the dependence of the switching exponent on $g$ is much weaker than the exponential dependence expected from Eq.~(\ref{eq:Asw_for_1D_Poisson}).

\begin{figure}[h]
 \includegraphics[clip,width=3in]{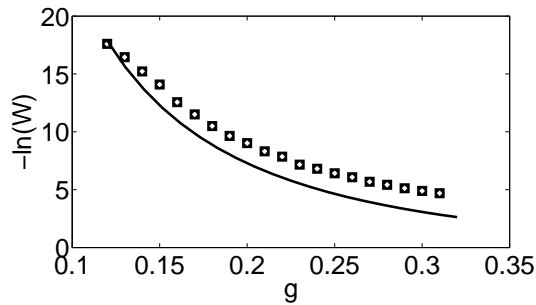}
\caption{Switching exponent $R_P=-\ln W$ for an overdamped particle in a
  potential (\ref{eq:quartic_potential}) driven by a Poisson noise with
  dimensionless mean frequency $\nu=0.5$ and pulse intensity $g$. Squares show
  the results of Monte-Carlo simulations, the solid line is the asymptotic theory (\ref{eq:switching_exponent_1D_Poisson}).}
\label{fig:Poisson_only}
\end{figure}

Numerical simulations of the switching rate for purely Poisson noise are shown in Fig.~\ref{fig:Poisson_only}. There is good agreement between the data and the asymptotic theory (\ref{eq:switching_exponent_1D_Poisson}) for small $g$. In this range $R_{P}g$ slowly varies with $g$.

In conclusion, we have considered switching in systems simultaneously driven by a Gaussian and a non-Gaussian noise. Even where the non-Gaussian noise has intensity smaller than that of the Gaussian, it may strongly change the switching rate. The effect is determined by the ratio of the non-Gaussian noise amplitude to the Gaussian noise intensity. It is described by the characteristic functional of the non-Gaussian noise calculated for a function determined by the system dynamics without this noise. A non-Gaussian tail of the noise distribution may strongly modify the switching rate even for a small noise intensity. We demonstrate this effect using Poisson noise as an example. Analytical results are compared with Monte Carlo simulations.

MID acknowledges valuable discussions with M. B\"uttiker and A. Korotkov. LB is supported by ARO grant No. W911NF-06-1-0320. MID is supported by ARO grant No.
W911NF-06-1-0324 and NSF grant  DMR-0305746. IBS is supported by the Office of Naval Research.


\begin{thebibliography}{33}
\expandafter\ifx\csname natexlab\endcsname\relax\def\natexlab#1{#1}\fi
\expandafter\ifx\csname bibnamefont\endcsname\relax
  \def\bibnamefont#1{#1}\fi
\expandafter\ifx\csname bibfnamefont\endcsname\relax
  \def\bibfnamefont#1{#1}\fi
\expandafter\ifx\csname citenamefont\endcsname\relax
  \def\citenamefont#1{#1}\fi
\expandafter\ifx\csname url\endcsname\relax
  \def\url#1{\texttt{#1}}\fi
\expandafter\ifx\csname urlprefix\endcsname\relax\def\urlprefix{URL }\fi
\providecommand{\bibinfo}[2]{#2}
\providecommand{\eprint}[2][]{\url{#2}}



\bibitem[{\citenamefont{Lapidus et~al.}(1999)\citenamefont{Lapidus, Enzer, and
  Gabrielse}}]{Lapidus1999}
\bibinfo{author}{\bibfnamefont{L.~J.} \bibnamefont{Lapidus}},
  \bibinfo{author}{\bibfnamefont{D.}~\bibnamefont{Enzer}}, \bibnamefont{and}
  \bibinfo{author}{\bibfnamefont{G.}~\bibnamefont{Gabrielse}},
  \bibinfo{journal}{Phys. Rev. Lett.} \textbf{\bibinfo{volume}{83}},
  \bibinfo{pages}{899} (\bibinfo{year}{1999}).

\bibitem[{\citenamefont{Siddiqi et~al.}(2005)\citenamefont{Siddiqi, Vijay,
  Pierre, Wilson, Frunzio, Metcalfe, Rigetti, Schoelkopf, Devoret, Vion
  et~al.}}]{Siddiqi2005}
\bibinfo{author}{\bibfnamefont{I.}~\bibnamefont{Siddiqi}}
 \bibnamefont{et~al.},
  \bibinfo{journal}{Phys. Rev. Lett.} \textbf{\bibinfo{volume}{94}},
  \bibinfo{pages}{027005} (\bibinfo{year}{2005}).

\bibitem[{\citenamefont{Aldridge and Cleland}(2005)}]{Aldridge2005}
\bibinfo{author}{\bibfnamefont{J.~S.} \bibnamefont{Aldridge}} \bibnamefont{and}
  \bibinfo{author}{\bibfnamefont{A.~N.} \bibnamefont{Cleland}},
  \bibinfo{journal}{Phys. Rev. Lett.} \textbf{\bibinfo{volume}{94}},
  \bibinfo{pages}{156403} (\bibinfo{year}{2005}).

\bibitem[{\citenamefont{Kim et~al.}(2005)\citenamefont{Kim, Heo, Lee, Ha, Jang,
  Noh, and Jhe}}]{Kim2005}
\bibinfo{author}{\bibfnamefont{K.}~\bibnamefont{Kim}}
   \bibnamefont{et~al.},
  \bibinfo{journal}{Phys. Rev. A} \textbf{\bibinfo{volume}{72}},
  \bibinfo{pages}{053402} (\bibinfo{year}{2005}).

\bibitem[{\citenamefont{Gommers et~al.}(2005)\citenamefont{Gommers, Douglas,
  Bergamini, Goonasekera, Jones, and Renzoni}}]{Gommers2005}
\bibinfo{author}{\bibfnamefont{R.}~\bibnamefont{Gommers}}
  \bibnamefont{et~al.},
  \bibinfo{journal}{Phys. Rev. Lett.} \textbf{\bibinfo{volume}{94}},
  \bibinfo{pages}{143001} (\bibinfo{year}{2005}).

\bibitem[{\citenamefont{Stambaugh and Chan}(2006)}]{Stambaugh2006}
\bibinfo{author}{\bibfnamefont{C.}~\bibnamefont{Stambaugh}} \bibnamefont{and}
  \bibinfo{author}{\bibfnamefont{H.~B.} \bibnamefont{Chan}},
  \bibinfo{journal}{Phys. Rev. B} \textbf{\bibinfo{volume}{73}},
  \bibinfo{pages}{172302} (\bibinfo{year}{2006}).

\bibitem[{\citenamefont{Abdo et~al.}(2007)\citenamefont{Abdo, Segev,
  Shtempluck, and Buks}}]{Abdo2007}
\bibinfo{author}{\bibfnamefont{B.}~\bibnamefont{Abdo}},
 \bibnamefont{et~al.},
  \bibinfo{journal}{J. Appl. Phys.} \textbf{\bibinfo{volume}{101}},
  \bibinfo{pages}{083909} (\bibinfo{year}{2007}).

\bibitem[{\citenamefont{Lupa\c{s}cu et~al.}(2007)\citenamefont{Lupa\c{s}cu,
  Saito, Picot, De~Groot, Harmans, and Mooij}}]{Lupascu2007}
\bibinfo{author}{\bibfnamefont{A.}~\bibnamefont{Lupa\c{s}cu}}
 \bibnamefont{et~al.},
  \bibinfo{journal}{Nature Physics} \textbf{\bibinfo{volume}{3}},
  \bibinfo{pages}{119} (\bibinfo{year}{2007}).

\bibitem[{\citenamefont{Katz et~al.}(2007)\citenamefont{Katz, Retzker, Straub,
  and Lifshitz}}]{Katz2007}
\bibinfo{author}{\bibfnamefont{I.}~\bibnamefont{Katz}}
  \bibnamefont{et~al.},
  \bibinfo{journal}{Phys. Rev. Lett.} \textbf{\bibinfo{volume}{99}},
  \bibinfo{pages}{040404} (\bibinfo{year}{2007}).

\bibitem[{\citenamefont{Serban and Wilhelm}(2007)}]{Serban2007}
\bibinfo{author}{\bibfnamefont{I.}~\bibnamefont{Serban}} \bibnamefont{and}
  \bibinfo{author}{\bibfnamefont{F.~K.} \bibnamefont{Wilhelm}},
  \bibinfo{journal}{Phys. Rev. Lett.} \textbf{\bibinfo{volume}{99}},
  \bibinfo{eid}{137001} (\bibinfo{year}{2007}).

\bibitem[{\citenamefont{Tobiska and Nazarov}(2004)}]{Tobiska2004}
\bibinfo{author}{\bibfnamefont{J.}~\bibnamefont{Tobiska}} \bibnamefont{and}
  \bibinfo{author}{\bibfnamefont{Y.}~\bibnamefont{Nazarov}},
  \bibinfo{journal}{Phys. Rev. Lett.} \textbf{\bibinfo{volume}{93}},
  \bibinfo{pages}{106801} (\bibinfo{year}{2004}).

\bibitem[{\citenamefont{Levitov and Lesovik}(1992)}]{Levitov1992}
\bibinfo{author}{\bibfnamefont{L.~S.} \bibnamefont{Levitov}} \bibnamefont{and}
  \bibinfo{author}{\bibfnamefont{G.~B.} \bibnamefont{Lesovik}},
  \bibinfo{journal}{JETP Lett.} \textbf{\bibinfo{volume}{55}},
  \bibinfo{pages}{555} (\bibinfo{year}{1992}).

\bibitem[{\citenamefont{Nazarov}(2007)}]{Nazarov2007}
\bibinfo{author}{\bibfnamefont{Y.~V.} \bibnamefont{Nazarov}},
  \bibinfo{journal}{Ann. Phys. (Leipzig)} \textbf{\bibinfo{volume}{16}},
  \bibinfo{pages}{720} (\bibinfo{year}{2007}).

\bibitem[{\citenamefont{Pekola}(2004)}]{Pekola2004}
\bibinfo{author}{\bibfnamefont{J.~P.} \bibnamefont{Pekola}},
  \bibinfo{journal}{Phys. Rev. Lett.} \textbf{\bibinfo{volume}{93}},
  \bibinfo{pages}{206601} (\bibinfo{year}{2004}).

\bibitem[{\citenamefont{Ankerhold}(2007)}]{Ankerhold2007a}
\bibinfo{author}{\bibfnamefont{J.}~\bibnamefont{Ankerhold}},
  \bibinfo{journal}{Phys. Rev. Lett.} \textbf{\bibinfo{volume}{98}},
  \bibinfo{pages}{036601} (\bibinfo{year}{2007}).

\bibitem[{\citenamefont{Sukhorukov and Jordan}(2007)}]{Sukhorukov2007}
\bibinfo{author}{\bibfnamefont{E.~V.} \bibnamefont{Sukhorukov}}
  \bibnamefont{and} \bibinfo{author}{\bibfnamefont{A.~N.}
  \bibnamefont{Jordan}}, \bibinfo{journal}{Phys. Rev. Lett.}
  \textbf{\bibinfo{volume}{98}}, \bibinfo{pages}{136803}
  (\bibinfo{year}{2007}).

\bibitem[{\citenamefont{Grabert}(2008)}]{Grabert2008}
\bibinfo{author}{\bibfnamefont{H.}~\bibnamefont{Grabert}},
  \bibinfo{journal}{Phys. Rev. B} \textbf{\bibinfo{volume}{77}},
  \bibinfo{pages}{205315} (\bibinfo{year}{2008}).

\bibitem[{\citenamefont{Timofeev et~al.}(2007)\citenamefont{Timofeev, Meschke,
  Peltonen, Heikkila, and Pekola}}]{Timofeev2007}
\bibinfo{author}{\bibfnamefont{A.~V.} \bibnamefont{Timofeev}}
 \bibnamefont{et~al.},
\bibinfo{journal}{Phys. Rev. Lett.}
  \textbf{\bibinfo{volume}{98}}, \bibinfo{pages}{207001}
  (\bibinfo{year}{2007}).

\bibitem[{\citenamefont{Huard et~al.}(2007)\citenamefont{Huard, Pothier, Birge,
  Esteve, Waintal, and Ankerhold}}]{Huard2007}
\bibinfo{author}{\bibfnamefont{B.}~\bibnamefont{Huard}}
  \bibnamefont{et~al.},
  \bibinfo{journal}{Ann. Phys. (Leipzig)} \textbf{\bibinfo{volume}{16}},
  \bibinfo{pages}{736} (\bibinfo{year}{2007}).

\bibitem[{\citenamefont{Novotn\'y}(2008)}]{Novotny2008}
\bibinfo{author}{\bibfnamefont{T.}~\bibnamefont{Novotn\'y}},
  \bibinfo{journal}{arxiv:0807.0387}  (\bibinfo{year}{2008}).

\bibitem[{\citenamefont{Sukhorukov and Jordan}(2008)}]{Sukhorukov2008}
\bibinfo{author}{\bibfnamefont{E.~V.} \bibnamefont{Sukhorukov}}
  \bibnamefont{and} \bibinfo{author}{\bibfnamefont{A.~N.}
  \bibnamefont{Jordan}}, \bibinfo{journal}{arXiv.org:0807.2675}
  (\bibinfo{year}{2008}).

\bibitem[{\citenamefont{Freidlin and Wentzell}(1998)}]{Freidlin_book}
\bibinfo{author}{\bibfnamefont{M.~I.} \bibnamefont{Freidlin}} \bibnamefont{and}
  \bibinfo{author}{\bibfnamefont{A.~D.} \bibnamefont{Wentzell}},
  \emph{\bibinfo{title}{Random Perturbations of Dynamical Systems}}
  (\bibinfo{publisher}{Springer-Verlag}, \bibinfo{address}{New York},
  \bibinfo{year}{1998}), \bibinfo{edition}{2nd} ed.

\bibitem[{\citenamefont{Smelyanskiy et~al.}(1997)\citenamefont{Smelyanskiy,
  Dykman, Rabitz, and Vugmeister}}]{Smelyanskiy1997c}
\bibinfo{author}{\bibfnamefont{V.~N.} \bibnamefont{Smelyanskiy}}
  \bibnamefont{et~al.},
  \bibinfo{journal}{Phys. Rev. Lett.} \textbf{\bibinfo{volume}{79}},
  \bibinfo{pages}{3113} (\bibinfo{year}{1997}).

\bibitem[{\citenamefont{Dykman et~al.}(1997)\citenamefont{Dykman, Rabitz,
  Smelyanskiy, and Vugmeister}}]{Dykman1997}
\bibinfo{author}{\bibfnamefont{M.~I.} \bibnamefont{Dykman}}
 \bibnamefont{et~al.},
  \textbf{\bibinfo{volume}{79}}, \bibinfo{pages}{1178} (\bibinfo{year}{1997}).

\bibitem[{\citenamefont{Kramers}(1940)}]{Kramers1940}
\bibinfo{author}{\bibfnamefont{H.}~\bibnamefont{Kramers}},
  \bibinfo{journal}{Physica (Utrecht)} \textbf{\bibinfo{volume}{7}},
  \bibinfo{pages}{284} (\bibinfo{year}{1940}).

\bibitem[{\citenamefont{Dykman}(1990)}]{Dykman1990}
\bibinfo{author}{\bibfnamefont{M.~I.} \bibnamefont{Dykman}},
  \bibinfo{journal}{Phys. Rev. A} \textbf{\bibinfo{volume}{42}},
  \bibinfo{pages}{2020} (\bibinfo{year}{1990}).

\bibitem[{\citenamefont{Dykman and Smelyanskiy}(1998)}]{Dykman1998b}
\bibinfo{author}{\bibfnamefont{M.~I.} \bibnamefont{Dykman}} \bibnamefont{and}
  \bibinfo{author}{\bibfnamefont{V.~N.} \bibnamefont{Smelyanskiy}},
  \bibinfo{journal}{Superlattices and Microstructures}
  \textbf{\bibinfo{volume}{23}}, \bibinfo{pages}{495} (\bibinfo{year}{1998}).

\bibitem[{\citenamefont{Assaf et~al.}(2008)\citenamefont{Assaf, Kamenev, and
  Meerson}}]{Assaf2008}
\bibinfo{author}{\bibfnamefont{M.}~\bibnamefont{Assaf}},
  \bibinfo{author}{\bibfnamefont{A.}~\bibnamefont{Kamenev}}, \bibnamefont{and}
  \bibinfo{author}{\bibfnamefont{B.}~\bibnamefont{Meerson}},
  \bibinfo{journal}{arxiv:0807.4812}  (\bibinfo{year}{2008}).

\bibitem[{\citenamefont{Dykman et~al.}(2008)\citenamefont{Dykman, Schwartz, and
  Landsman}}]{Dykman2008}
\bibinfo{author}{\bibfnamefont{M.~I.} \bibnamefont{Dykman}},
  \bibinfo{author}{\bibfnamefont{I.~B.} \bibnamefont{Schwartz}},
  \bibnamefont{and} \bibinfo{author}{\bibfnamefont{A.~S.}
  \bibnamefont{Landsman}}, \bibinfo{journal}{Phys. Rev. Lett.}
  \textbf{\bibinfo{volume}{101}}, \bibinfo{pages}{078101}
  (\bibinfo{year}{2008}).

\bibitem[{\citenamefont{Feynman and Hibbs}(1965)}]{FeynmanQM}
\bibinfo{author}{\bibfnamefont{R.~P.} \bibnamefont{Feynman}} \bibnamefont{and}
  \bibinfo{author}{\bibfnamefont{A.~R.} \bibnamefont{Hibbs}},
  \emph{\bibinfo{title}{Quantum Mechanics and Path Integrals}}
  (\bibinfo{publisher}{McGraw-Hill}, \bibinfo{address}{New-York},
  \bibinfo{year}{1965}).

\bibitem[{\citenamefont{McKane}(1989)}]{Mckane1989}
\bibinfo{author}{\bibfnamefont{A.~J.} \bibnamefont{McKane}},
  \bibinfo{journal}{Phys. Rev. A} \textbf{\bibinfo{volume}{40}},
  \bibinfo{pages}{4050} (\bibinfo{year}{1989}).

\bibitem[{\citenamefont{Luciani and Verga}(1987)}]{Luciani1987}
\bibinfo{author}{\bibfnamefont{J.}~\bibnamefont{Luciani}} \bibnamefont{and}
  \bibinfo{author}{\bibfnamefont{A.}~\bibnamefont{Verga}},
  \bibinfo{journal}{Europhys. Lett.} \textbf{\bibinfo{volume}{4}},
  \bibinfo{pages}{255} (\bibinfo{year}{1987}).

\bibitem[{\citenamefont{Cohen}()}]{Cohen_private}
\bibinfo{author}{\bibfnamefont{E.~G.~D.} \bibnamefont{Cohen}},
  \bibinfo{howpublished}{private communication}.

\end{thebibliography}

\end{document}